\documentstyle[prd,twocolumn,aps,epsf]{revtex}
\newcommand{\singlefig}[2]{
\begin{center}
\begin{minipage}{#1}
\epsfxsize=#1
\epsffile{#2}
\end{minipage}
\end{center}}
\newcommand{\segmentfig}[3]{
\begin{minipage}{#1}
\epsfxsize=#1
\epsffile{#2}
\begin{center}
{\small \mbox{#3}}
\end{center}
\end{minipage}}
\newcommand{\gsim}{\mbox{\raisebox{-1.ex}{$\stackrel
     {\textstyle>}{\textstyle\sim}$}}}
\newcommand{\lsim}{\mbox{\raisebox{-1.ex}{$\stackrel
     {\textstyle<}{\textstyle \sim}$}}}
\begin{document}
\draft
\title{Estimating Hawking radiation for exotic black holes}
\author{Takashi Tamaki\thanks{electronic
mail:tamaki@gravity.phys.waseda.ac.jp}
}
\address{Department of Physics, Waseda University,
Shinjuku, Tokyo 169-8555, Japan}
\author{Kei-ichi 
Maeda\thanks{electronic mail:maeda@gravity.phys.waseda.ac.jp} 
}
\address{Department of Physics, Waseda University,
Shinjuku, Tokyo 169-8555, Japan}
\address{Advanced Research Institute for Science and Engineering, 
Waseda University, Shinjuku, Tokyo 169-8555, Japan}
\date{\today}
\maketitle
\begin{abstract}
We study about an approximation method of the Hawking radiation. 
We analyze an massless scalar field 
in exotic black hole backgrounds
models which have peculiar properties in black hole thermodynamics 
(monopole black hole in SO(3) Einstein-Yang-Mills-Higgs system 
and dilatoic black hole 
in Einstein-Maxwell-dilaton system). 
A scalar field is assumed not to be couple to matter fields consisting 
of a black hole background. 
Except for extreme black holes, we can well approximate the Hawking radiaition 
by `black body' one with Hawking temperature estimated at a radius 
of a critical impact parameter. 
\end{abstract}

The Hawking radiaition\cite{Hawking} around black holes has been discussed 
for many years concerning various aspects, e.g., as $\gamma$-ray sources 
of the early universe\cite{Page} or vacuum polarization of charged black holes
\cite{Zaumen}, etc. But as for black holes with non-Abelian hair
\cite{Tor2,colored,Torii,DHS,Greene,LM,Tor}, 
it has been less investigated because many of them are only obtained 
numerically that it takes much works compared with ones of analytically obtained. 
But we need to investigate for many reasons. Particularly, a monopole black hole 
which was found in SO(3) Einstein-Yang-Mill-Higgs (EYMH) system
\cite{BFM,Lee,Aichelberg,Tachi} is important because 
it is one of the counterexample of the black hole no hair conjecture\cite{Bizon}.
Moreover, if we consider the evaporation process of the Reissner-Nortstr\"om (RN) 
black hole, it may become a monopole black hole and the final product (a regular 
gravitating monopole) could be a good candidate for the remnant of the 
Hawking radiation in such a system. 

From such view points, we study an approximation method of the Hawking radiation 
which may reduce our labor and 
discuss its validity in this paper. 
Throughout this paper we use 
units $c=\hbar =1$. Notations and definitions as such as Christoffel symbols 
and curvature follow Misner-Thorne-Wheeler\cite{MTW}.


We consider two models in which black hole solutions have peculiar properties in 
black hole thermodynamics. 

(I) The SO(3) EYMH model as
\begin{eqnarray}
S=\int d^{4}x\sqrt{-g}\left[\frac{R}{2\kappa^{2}} + L_{m}\right],
\label{EYMH}
\end{eqnarray}
where $\kappa^{2} \equiv 8\pi G$ with $G$ being Newton's
gravitational constant. $L_{m}$ is  the Lagrangian density of the matter fields which 
are written as 
\begin{equation}
L_{m}  =  -\frac{1}{4}F^{a}_{\mu\nu}F^{a\mu\nu}
-\frac{1}{2}(  D_{\mu}\Phi^{a}  )(  D^{\mu}\Phi^{a}  )
-\frac{\lambda}{4}(\Phi^{a}\Phi^{a}-v^{2})^{2}\ .
\label{matter lag}     
\end{equation}
$F^{a}_{\mu\nu}$ is the field strength of the SU(2) YM field 
and expressed by its potential $A^{a}_{\mu}$ as
\begin{equation}
F^{a}_{\mu\nu} =  \partial_{\mu}A^{a}_{\mu}-
\partial_{\nu}A^{a}_{\nu}+e \epsilon^{abc}A^{b}_{\mu}A^{c}_{\nu},
\label{B6}     
\end{equation}
with a gauge coupling constant $e$.
$\Phi^{a}$ is a 
real triplet Higgs field and $D_{\mu}$ is the covariant derivative: 
\begin{equation}
D_{\mu}\Phi^{a} =  \partial_{\mu} \Phi^{a}+
e \epsilon^{abc}A^{b}_{\mu}\Phi^{c}.
\label{B7}
\end{equation}
The parameters $v$ and $\lambda$ are the 
vacuum expectation value and a self-coupling 
constant of the Higgs field, respectively. 

(II) The Einstein-Maxwell-dilaton (EMD) model as 
\begin{eqnarray}
S=\frac{1}{2\kappa^{2}}\int d^{4}x 
\sqrt{-g}\left[R-2(\nabla \phi)^{2}-e^{-2\alpha \phi} F^{2}\right], 
\label{dilaton} 
\end{eqnarray}
where $\phi$ and $F$ are a dilaton field and U(1) gauge field, respectively. 

For black hole solutions, 
we assume that a space-time is static and spherically symmetric, in which case 
the metric is written as 
\begin{equation}
ds^{2}=-f(r)e^{-2\delta (r)}dt^{2}+
f(r)^{-1}dr^{2}+r^{2}d\Omega^{2},
\label{B8}
\end{equation}
where $f(r)=1-2Gm(r)/r$. 
We consider spacetime, which have a regular horizon and is asymptotically flat.  
In such a background spacetime, we consider a neutral and massless scalar field which does not 
couple to the matter fields such as the YM field or the Higgs field. 
The eq. of motion is described by the Klein-Gordon 
equation as 
\begin{eqnarray}
\Phi_{,\mu}^{\ ;\mu}=0 .
\label{Klein-Gordon} 
\end{eqnarray}
This equation is separable, 
and we should only solve the radial equation 
\begin{eqnarray}
\frac{ d^{2}\chi }{ d{r}^{\ast 2} }+\chi\left[ \omega^{2}
-V^{2} \right]=0 , 
\label{radial} 
\end{eqnarray}
where 
\begin{eqnarray}
V^{2}&\equiv &\frac{f}{re^{2\delta}}
\left\{ \frac{l(l+1)}{ r}-f\delta^{\prime}+
\frac{2(m-m^{\prime}r)}{r^{2}}
 \right\}  , 
\label{potential}  \\
\frac{dr}{dr^{\ast}}&\equiv & f e^{-\delta} ,
\label{tortus}
\end{eqnarray}
where $^{\prime}$ denote $d/dr$ and $\chi$ is only the function of $r$.  

The energy emission rate of Hawking radiation is given by 
\begin{eqnarray}
\Xi=\frac{1}{2\pi}\sum_{l=0}^{\infty}(2l+1)
\int_{0}^{\infty}\frac{  \Gamma(\omega)\omega  }{  
e^{\omega/T_{H}}-1  }d\omega ,
\label{emission} 
\end{eqnarray}
where $l$ and $\Gamma(\omega)$ are the angular momentum and 
the transmission probability in a scattering problem for the scalar 
field $\Phi$. $\omega$ and $T_{H}$ are the energy of 
the particle and the Hawking temperature 
respectively. Note that $\Xi \equiv -dM/dt$.  

The transmission probability $\Gamma$ can be calculated by 
solving radial equation (\ref{radial}) numerically 
under the boundary condition 
\begin{eqnarray}
\chi \rightarrow Ae^{-i\omega r^{\ast}}+Be^{i\omega r^{\ast}}
\ \ (r^{\ast} \rightarrow \infty ) ,
\label{infty}  \\
\chi \rightarrow e^{-i\omega r^{\ast}}
\ \ (r^{\ast} \rightarrow -\infty ) ,
\label{horizon} 
\end{eqnarray}
where $\Gamma$ is given as  $1/|A|^{2}$. As we see the dominant contribution 
for the Hawking radiation is $l=0$, because the contribution of the higher 
modes are suppressed by the centrifugal barrier. In what follows, we ignore the 
contributions from $l \geq 2$. Since the systematic 
error caused by it is below $1 \%$, 
this does not affect the discussion below. The evaporation process 
of the monopole black hole and the dilatonic black holes were discussed 
numerically in \cite{Tama} and in \cite{JK}, respectively.

In this note, we study an approximation method of the Hawking radiation 
and check how accurate it is. 
Naively speaking, the Hawking radiation is a blackbody radiation with the 
temperature $T_{H}$. Then we may believe that $T_{H}$ mostly 
contributes to the Hawking radiation and determines the 
evaporation rate of a black hole. 
However we have another factor, i.e., transmission amplitude $\Gamma$. 
Since $\Gamma$ depends on the energy of the particle, 
we have to solve a radial equation and integrate 
(\ref{emission}). Moreover, 
we have to solve the background spacetime numerically 
for non-Abelian black holes. It would be rather a troublesome task. 

But since the emission is of a blackbody nature, we may be able to estimate it by  
Stefan's law\cite{Wald} as $\Xi_{BB} \equiv \pi^{3} 
R_{eff}^{2}T_{H}^{4}/30$. 
The unknown is the effective radius $R_{eff}$. 
It has been suggested that for a Schwarzschild black hole, 
$R_{eff}$ is given by the photon orbit with some energy $E$ and 
angular momentum $L\equiv g_{ab}\Psi^{a}u^{b}$, where 
$\Psi^{a}$ and $u^{b}$ is the rotational Killing field and 
the 4-velocity, respectively. 
Although such a formula provides a good approximation 
in a Schwarzschild background, one may wonder whether 
it is still valid for exotic black holes, 
which have an envelope outside of the event horizon. 
 
Starting with metric (\ref{B8}) and considering a massless 
particle in such a background, 
we find the maximum point of the effective potential $V_{eff}$ defined by
\begin{eqnarray}
V_{eff}\equiv \frac{L^{2}}{2r^{2}}\left(1-\frac{2Gm(r)}{r}\right) ,
\label{efpotential} 
\end{eqnarray}
and then the effective radius as 
\begin{eqnarray}
R_{eff}\equiv \frac{L}{E}=\left( \frac{R^{3}}{R-2m(R)} \right)^{1/2}
e^{\delta (R)}, 
\label{efradius} 
\end{eqnarray}
where $R$ is the radial cordinate where $V_{eff}$ takes its maximum value. 
$R_{eff}$ is determined by the 
critical value of the ``impact parameter" below which 
any ``photon" sent toward the black hole 
can not escape. 
Using this effective radius, 
we define $\Xi_{BB} \equiv \pi^{3}R_{eff}^{2}T_{H}^{4}/30$ 
as a blackbody approximation. 

First, we examine how well we can approximate by this method in monopole and 
RN black holes. We show a relation between the Hawking temperature $T_{H}$ 
and the ratio $\Xi /\Xi_{BB}$ in Fig. 1. 
We choose as $\lambda /e^{2}=0.1$, $v/M_{pl}=0.05$. The arrow in this diagram 
shows the direction of the smaller mass, 
which corresponds to that of the evaporating process. 
We find a quick turn below the point $A$ which corresponds to 
the sign change of the specific heat. Our 
calculation ends at $Q/Q_{max}=0.99$ where $Q_{max}$ 
is a maximum charge.  
As for other parameter choice of $\lambda$ and $v$, 
we obtain similar results. 
We summerize the results as follows. 

(i) For more generic cases, the approximation 
is very good, but it becomes wrong near the extreme limit. 

(ii) The temperature is not a good measure for the 
validity of this approximation. For example, 
if we have two solutions with the same temperature as in Fig. 1, 
a near-extreme solution gives worse result. 

What condition should be imposed for good approximation? 
To clarify this, we consider the EMD model 
in which exact black hole solutions are obtained\cite{GM}.
One of the reason to choose this model is 
the thermodynamical properties of the black hole 
depends on the coupling $\alpha$. 
Particularly, $\alpha$ changes the temperature and the shapes of the effective 
potential near the extreme solution. Roughly speaking, 
we may think its validity by 
two factors, steepness of the effective potential and the temperature 
which determines the frequency of 
particles which contribute to the emission rate.

We show shapes of the 
potentials for (a) $\alpha =0$ (b) $\alpha =1$ and 
(c) $\alpha =2$ in terms of $r^{\ast}$ in Fig. 2\cite{footnote1}. 
The parameters $Q/Q_{max}$ are chosen to be $0$, $0.84$, $0.89$, $0.94$, $0.99$ 
in (a) and (b) and to be $0$, $0.81$, $0.86$, $0.9$, $0.95$ in (c), 
respectively. 
As we increase $\alpha$, the difference from 
the Schwarzschild black hole increases 
near the extreme limit, i.e., the shapes of the potential becomes 
steep and the temperature becomes high. So its a 
competition between steepness of the potential which 
makes approximation wrong and the increasing 
temperature which improves the approximation.  

We show the relation between $Q/Q_{max}$ and $\Xi /\Xi_{BB}$ for $\alpha =0$, $0.5$, 
$1.0$, $1.5$, $2.0$ in Fig. 3. For $Q/Q_{max}\lsim 0.8$, 
$\Xi /\Xi_{BB}-1$ is about $0.06$, and it seems to be 
rather universal. But for $Q/Q_{max}\gsim 0.8$, results depend on $\alpha$ 
remarkably. 
It may seem strange because the value $\Xi/\Xi_{BB}$ near the extreme solution 
does not monotonically increase by the increase of $\alpha$ 
(The order for `good' approximation is $\alpha=2.0$, $1.5$, $0.0$, $1.0$, $0.5$.), 
but we can interpret it easily. 

We show the relation between $T_{H}$ and $\Xi /\Xi_{BB}$ 
in Fig. 4 for the same parameters in Fig. 3. The arrows show the direction 
to the evaporation process. If we compare the $\Xi /\Xi_{BB}$ for fixed $T_{H}$, 
it monotonically depends on the `steepness' of the potential which can be 
evaluated using $R_{eff}$ because it decreases when $V_{eff}$ becomes steep. 
Moreover, $R_{eff}$ can be the criterion for evaluating the curvature radius 
of the spacetime. Since $T_{H}$ is the good indicator of the mean frequency, 
to see $T_{H}R_{eff}$ would reveal the validity of the geometric optics 
approximation as was already pointed out in \cite{Wald}. 

We show $T_{H}R_{eff}$ in terms of 
$Q/Q_{max}$ in Fig. 5. One can see that $T_{H}R_{eff}$ resembles to the 
$\Xi /\Xi_{BB}$. For the geometric optics approximation to be justified, 
it is usually demanded that $T_{H}R_{eff}$ should be far larger than $1$. 
But this figure shows that $\Xi_{BB}$ could work even in the case 
$T_{H}R_{eff}\sim 0.2$. It is surprising that this crude approximation 
provide such good results and it 
may provide the effective way to evaluate the Hawking radiation.

\section*{ACKOWLEDGEMENTS}
Special Thanks to J. Koga, T. Torii and T. Tachizawa for useful discussions. 
T. T is thankful for  financial support from the JSPS (No. 106613).


\begin{figure}
\begin{center}
\singlefig{8cm}{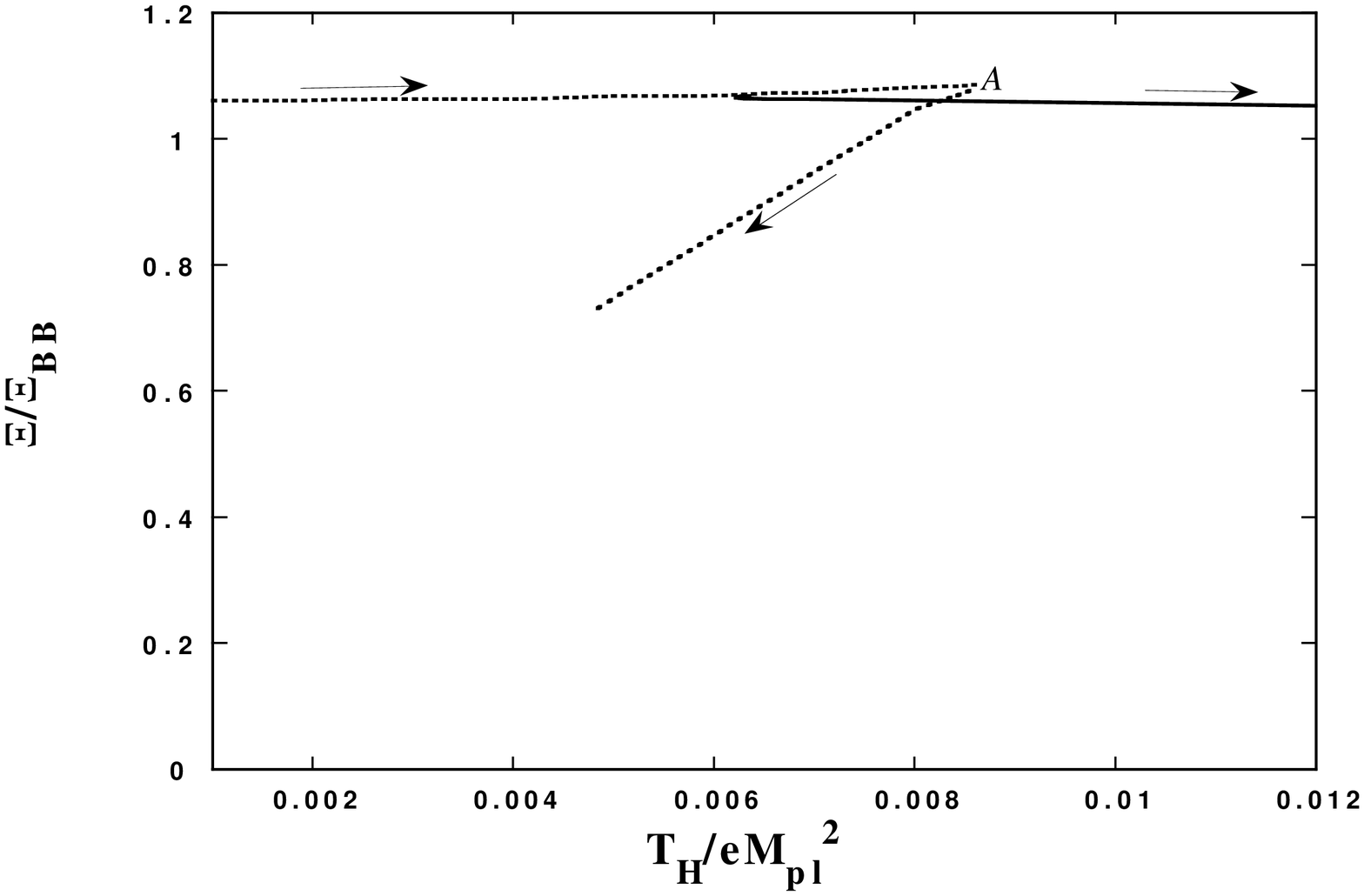}
\caption{The ratio between the Hawking radiation and the black body 
approximation $\Xi /\Xi_{BB}$ 
in terms of the Hawking temperature $T_{H}/eM_{pl}^{2}$ 
for $\lambda /e^{2}=0.1$, $v/M_{pl}=0.05$. We show the RN black hole as dotted lines 
and the monopole black hole as solid lines. The point $A$ correspond to the change 
of the sign of the specific heat. We can find that though 
$\Xi_{BB}$ differs from $\Xi$ about $6\%$, its difference seems rather universal 
except below the point $A$.  
\label{kinji-mp} }
\end{center}
\end{figure}
\begin{figure}[htbp]
\segmentfig{8cm}{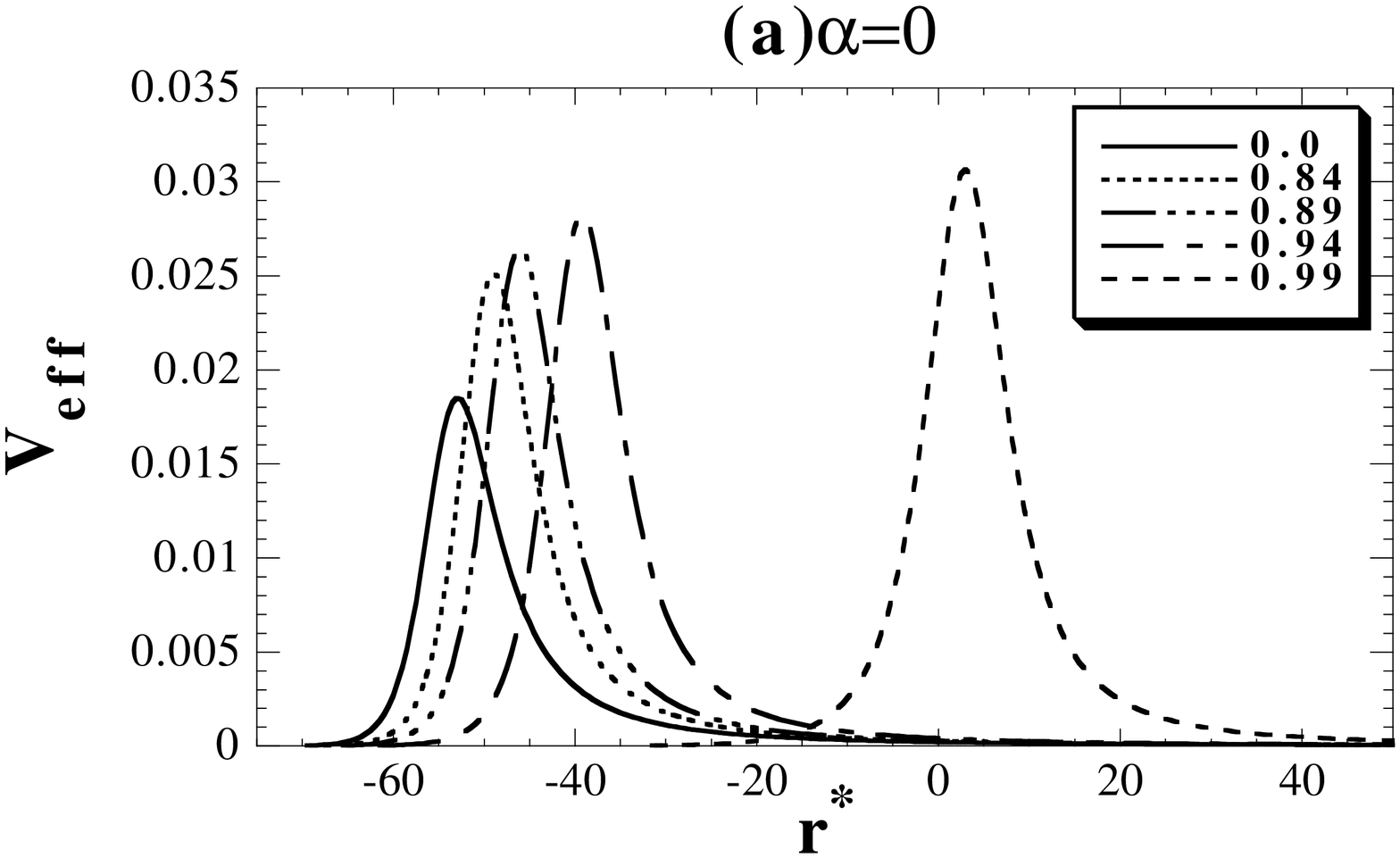}{}  \\
\segmentfig{8cm}{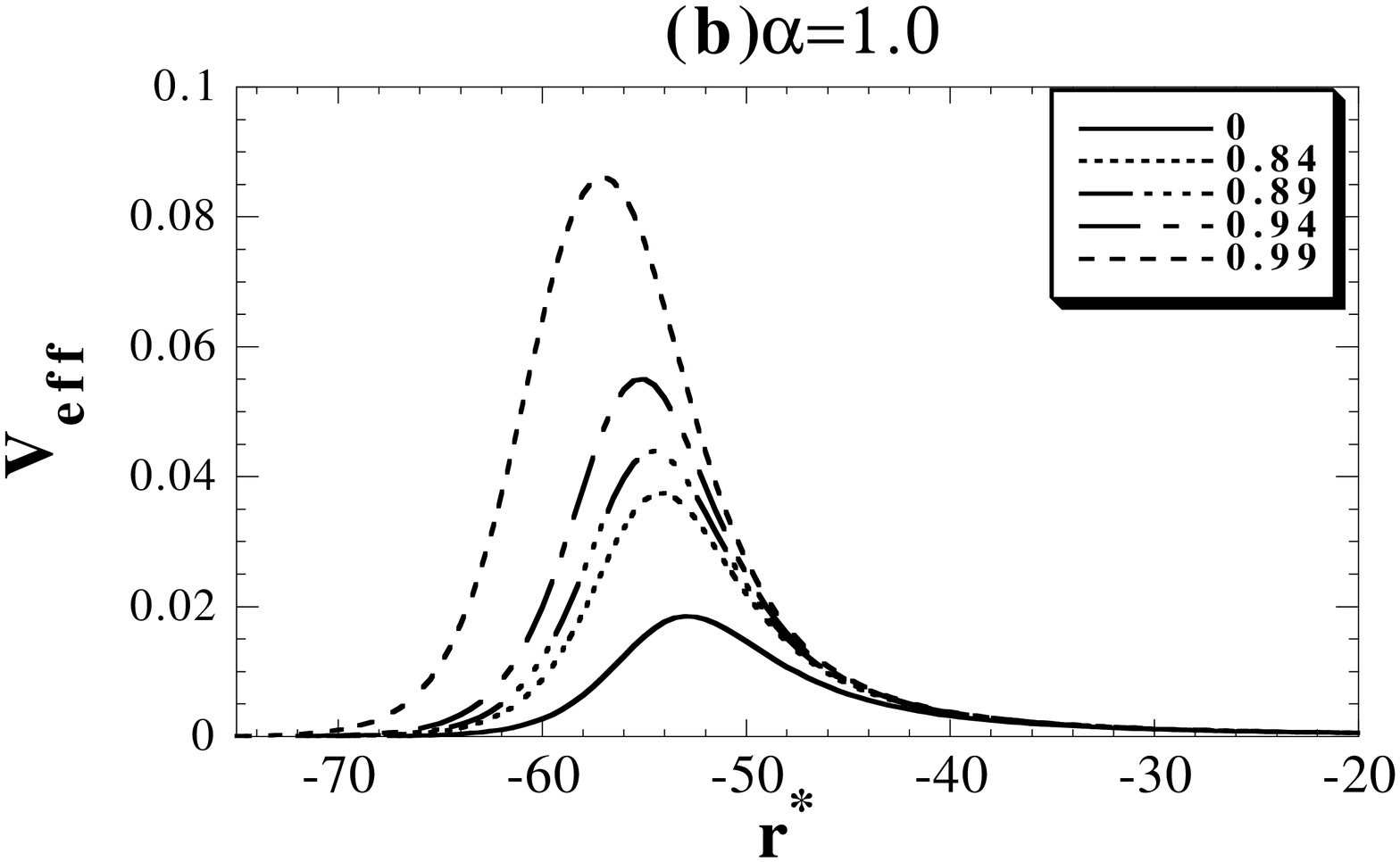}{}  \\
\segmentfig{8cm}{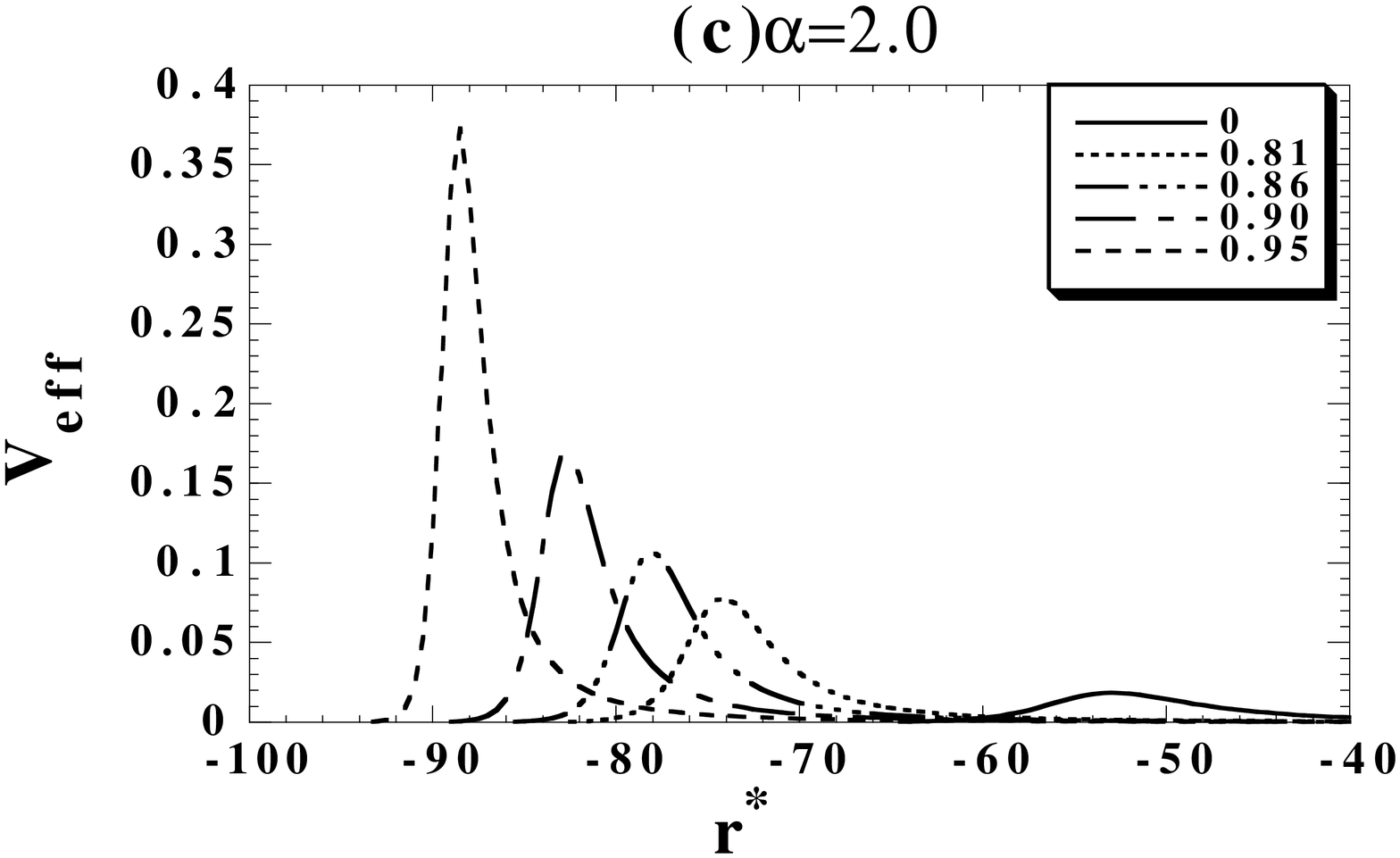}{}  
\caption{Shapes of the potential $V_{eff}$ in terms of $r^{\ast}$ 
for (a) $\alpha =0$ 
(b) $\alpha =1$ (c) $\alpha =2$. 
For the parameter $Q/Q_{max}$, we choose to be 
$0$, $0.84$, $0.89$, $0.94$, $0.99$ in (a) and (b) and to be 
$0$, $0.81$, $0.86$, $0.9$, $0.95$ in (c). 
Note that for large $\alpha$, the shapes of the 
potential extremely change near the extreme limit. 
\label{rast-veff}
}
\end{figure}
\begin{figure}
\begin{center}
\singlefig{7cm}{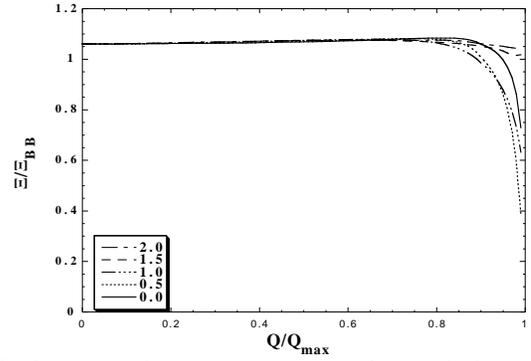}
\caption{The ratio between the Hawking radiation and 
the black body approximation $\Xi /\Xi_{BB}$ 
in terms of $Q/Q_{max}$ for dilatonic black holes for $\alpha =0$, $0.5$, $1$, 
$1.5$, $2$ which shows that near the extremal limit, $\Xi/\Xi_{BB}$ change rapidly. 
\label{kinji-dilaton} }
\end{center}
\end{figure}
\begin{figure}
\begin{center}
\singlefig{7cm}{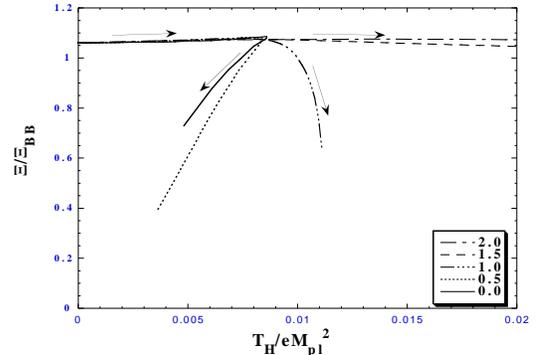}
\caption{The ratio between the Hawking radiation and 
the black body approximation $\Xi /\Xi_{BB}$ 
in terms of Hawking temperature $T_{H}/eM_{pl}^{2}$ for the same solutions in Fig. 3. 
The arrows show 
the decrease of the gravitational mass if the electric charge is fixed. 
This may suggests that the approximation becomes invalid 
for the decrease of $T_{H}$ or approaching the extreme black hole. 
\label{kinji-jyouken} }
\end{center}
\end{figure}
\begin{figure}
\begin{center}
\singlefig{7cm}{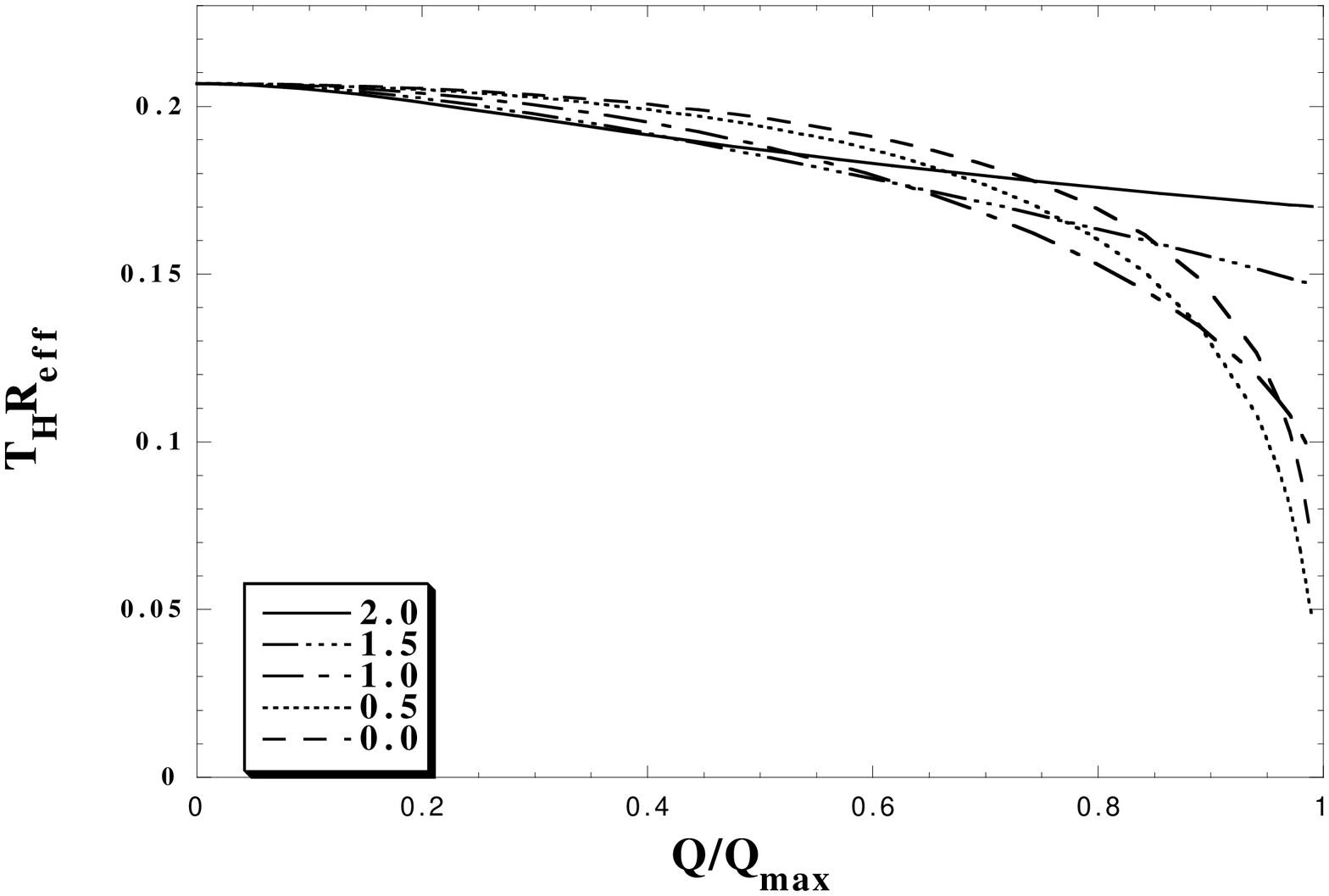}
\caption{$T_{H}R_{eff}$ in terms of $Q/Q_{max}$ which shows that this can 
be the indicator to evaluate the validity of the approximation. 
\label{kinji-jyouken2} }
\end{center}
\end{figure}


\begin{thebibliography}{99}
\bibitem{Hawking}
S. W. Hawking, Nature {\bf 248}, 30 (1974); Commun. Math. Phys. {\bf 43}, 
199 (1975). 
\bibitem{Page}
D. N. Page and S. W. Hawking, Astrophys. J. {\bf 206}, 1 (1976). 
\bibitem{Zaumen}
W. T. Zaumen, Nature {\bf 247}, 530 (1974); 
G. W. Gibbons, Comm. Math. Phys. {\bf 44}, 245 (1975). 
\bibitem{Tor2} As a review paper see K. Maeda, Journal of the 
Korean Phys. Soc. {\bf 28}, S468, (1995), 
and M. S. Volkov and D. V. Gal'tsov, hep-th/9810070.
\bibitem{colored}
M. S. Volkov and D. V. Galt'sov, Pis'ma Zh. Eksp. Theor. Fiz. {\bf 50}, 
312 (1989); P. Bizon, Phys. Rev. Lett. {\bf 64}, 2844 (1990); H. P. K\"
{u}nzle 
and A. K. Masoud-ul-Alam, J. Math. Phys. {\bf 31}, 928 (1990).
\bibitem{Torii}
K. Maeda, T. Tachizawa, T. Torii and T. Maki, Phys. Rev.
Lett. {\bf 72}, 450 (1994);
T. Torii, K. Maeda and T. Tachizawa, Phys. Rev. D. {\bf 51},
1510 (1995).
\bibitem{DHS}
S. Droz, M. Heusler and N. Straumann, 
Phys. Lett. B {\bf 268}, 371 (1991).
\bibitem{Greene}
B. R. Greene, S. D. Mathur and C. M. O'Neill, Phys. Rev. D {\bf 47}, 
2242 (1993).
\bibitem{LM}
H. Luckock and I. G. Moss, Phys. Lett. B {\bf 176}, 341 (1986); 
H. Luckock, in {\it String theory, quantum cosmology and quantum 
gravity, integrable and conformal invariant theories}, eds. H. de Vega 
and N. Sanchez, (World Scientific, Singapore, 1986), p. 455. 
\bibitem{Tor} 
T. Torii and K. Maeda, Phys. Rev. D {\bf 48}, 1643 (1993).
\bibitem{BFM}
P. Breitenlohner, P. Forg\'{a}cs and D. Maison, Nucl. Phys. B 
{\bf 383}, 357 (1992);  {\it ibid.} {\bf 442}, 126 (1995).
\bibitem{Lee}
K. -Y. Lee, V. P. Nair and E. Weinberg, Phys. Rev. Lett. {\bf 68}, 1100 
(1992); Phys. Rev. D {\bf 45}, 2751 (1992); Gen. Relativ. Gravit. 
{\bf 24}, 1203 (1992). 
\bibitem{Aichelberg}
P. C. Aichelberg and P. Bizon, Phys. Rev. D. {\bf 48}, 607 (1993). 
\bibitem{Tachi}
T. Tachizawa, K. Maeda  and T. Torii, Phys. Rev. D.
{\bf 51},  4054 (1995). 
\bibitem{Bizon}
P. Bizon, Acta. Phys. Pol. B {\bf 25}, 877 (1994). 
\bibitem{MTW}
C. W. Misner, K. S. Thorne and J. A. Wheeler, {\it
Gravitation} (Freeman, New York 1973). 
\bibitem{Tama}
T. Tamaki and K. Maeda, gr-qc/9910024. 
\bibitem{JK}
J. Koga and K. Maeda,  Phys. Rev. D. {\bf 52}, 7066 (1995). 
\bibitem{Wald}
R. M. Wald, {\it General\ Relativity} (Chicago Press, Chicago and London 1984). 
\bibitem{GM}
G.W. Gibbons and K. Maeda, Nucl. Phys. B {\bf 298}, 741 (1988); 
D. Garfinkle, G. T. Horowitz and A. Strominger, 
Phys. Rev. D {\bf 43}, 3140 (1991). 
\bibitem{footnote1}
We use the same notation as in \protect\cite{JK}. 
\end{thebibliography}
\end{document}